\documentclass[10pt]{iopart}

\usepackage{amssymb}
\usepackage{graphicx}
\usepackage{dcolumn}
\usepackage{bm}
\usepackage{graphicx, array}
\usepackage{rotating}
\usepackage[caption=false]{subfig}

\newcommand{\GRRU}{Gr/Ru(0001)}
\newcommand{\GRNI}{Gr/Ni(111)}

\begin{document}

\title{Effect of moir\'e superlattice reconstruction in the electronic excitation spectrum of graphene-metal heterostructures}

\author{Antonio Politano$^1$\footnote{Contributed equally to this work}, Guus J. Slotman$^2\ddagger$, Rafael Rold\'an$^3$, Gennaro Chiarello$^1$, Davide Campi$^4$, Mikhail I. Katsnelson$^2$, Shengjun Yuan$^2$}

\address{$^1$Department of
Physics, Universit$\grave{a}$ della Calabria, I-87036 Rende,Italy}
\address{$^2$Institute for Molecules and Materials, Radboud University, Heyendaalseweg 135, 6525AJ Nijmegen, The Netherlands}
\address{$^3$Instituto de Ciencia de Materiales de Madrid, CSIC, Cantoblanco, E-28049 Madrid, Spain}
\address{$^4$Department of Materials Science, University Milano-Bicocca, I-20125, Milano, Italy}
\ead{antonio.politano@fis.unical.it,  rroldan@icmm.csic.es, s.yuan@science.ru.nl}

\date{\today}

\begin{abstract}
We have studied the electronic excitation spectrum in periodically
rippled graphene on Ru(0001) and flat, commensurate graphene on Ni(111)
by means of high-resolution electron energy loss spectroscopy and
a combination of density functional theory and tight-binding approaches. We
show that the periodic moir\'e superlattice originated by the lattice mismatch in
graphene/Ru(0001) induces the emergence of an extra mode, which is not present in
graphene/Ni(111). Contrary to the ordinary intra-band plasmon of doped graphene,
the extra mode is robust in charge-neutral graphene/metal contacts, having its origin in
electron-hole inter-band transitions between van Hove singularities that emerge
in the reconstructed band structure, due to the moir\'e pattern superlattice.
\end{abstract}


\submitto{\TDM}

\maketitle
\ioptwocol

Large-scale, highly crystalline graphene can be grown on metal
substrates by chemical vapour deposition (CVD).\cite{Batzill201283} Depending on lattice mismatch between
graphene (Gr) and the underlying substrate, graphene can grow as
commensurate or incommensurate sheets. In the latter case, a
moir\'e superlattice emerges \cite{pletikosic2009dirac}. The
moir\'e-derived superperiodic potential leads to a reconstruction
of the electronic band structure, with emergence of mini-bands and
Dirac-cone replicas in the moir\'e superlattice Brillouin zone,
leading to a wealth of intriguing
phenomena~\cite{pletikosic2009dirac, xu2011exploring,
voloshina2013electronic,Wallbank2013,SanJose2014}. The spatial
modulation in periodically rippled graphene determines the
existence of localized electronic states in the high and low areas
of the moir\'e structure at energies close to the Fermi
level~\cite{stradi2012electron}. The band structure reconstruction
imposed by the periodic moir\'e potential implies a modification
of screening properties and, thus, of the plasmonic excitations
\cite{TGP14}. Inter-band transitions associated with the
superlattice mini-bands contribute to the excitation spectrum of
quasi-periodic Gr/hBN superlattices, as recently shown by
infrared nano-imaging experiments \cite{Ni2015}. Similar effects associated with the moir\'e pattern  have been observed in the phonon spectra of \GRRU~and \GRNI\cite{maccariello2015low,maccariello2015observation,al2016acoustic}. Whereas optical experiments typically probe modes of small wave-vectors, of the order of the photon momentum, spectroscopies based on inelastic scattering of impinging electrons allow exploring the excitation spectrum along the whole Brillouin zone.

Herein, we take as model systems periodically rippled \GRRU~and flat, commensurate graphene on \GRNI~to put in
evidence the effects of the moir\'e reconstruction on plasmonic
excitations in graphene. For this aim we combine high-resolution
electron energy loss spectroscopy (HREELS) with theoretical
\textit{ab initio} density functional theory (DFT) and tight-binding
propagation method (TBPM) calculations \cite{yuan2010modeling,hams2000fast}. Due to the lattice
mismatch with the Ru lattice, graphene forms a moir\'e
reconstruction on Ru(0001)~\cite{wang2008chemical,
borca2010electronic,Iannuzzi_2013}, well approximated as a ($12\times12$). The
geometric corrugation of graphene nanodomes on Ru(0001) has been
imaged by scanning tunneling microscopy
(STM)~\cite{borca2010electronic, marchini2007scanning}.
Periodically inhomogeneous electronic properties in valleys and
hills of the ripples have been inferred by means of scanning
tunneling spectroscopy (STS)~\cite{gyamfi2011inhomogeneous}, which
shows the occurrence of (i) localized
states~\cite{stradi2012electron}, (ii) splitting of image
states~\cite{borca2010potential} and (iii) local variations of the
work function~\cite{feng2011periodically}. Peaks observed in STS
near the Fermi level arise from localized electronic states in the
hills of the moir\'e pattern~\cite{stradi2012electron}.
Conversely, the close lattice match between graphene and Ni allows
the growth of a commensurate graphene overlayer on
Ni(111)~\cite{shikin1999surface,gamo1997atomic,dahal2014graphene}. Despite the hybridization of
Ni-$d_{3z^2-r^2}$ orbitals with $\pi$ states of graphene~\cite{Andersen2012}, the adsorption energy still remains in the
range of a typical physisorption (67 meV per carbon
atom)~\cite{mittendorfer2011graphene}. We find that commensurate
graphene on Ni(111) exhibits a single plasmon peak, associated to well known
standard intra-band collective charge oscillation with dispersion
$\hbar\omega_{pl}\propto \sqrt{q_{\parallel}}$, in terms of the in-plane wave-vector $q_{\parallel}$. On the other hand, \GRRU~shows (besides the standard intra-band plasmon) an
additional gapped and weakly dispersing mode in the spectrum. Our
TBPM calculations, based on structures determined by atomistic simulations,
suggest that the extra mode is associated to inter-band
particle-hole excitations between van Hove singularities generated
in the reconstructed band structure of the moir\'e superlattice.

HREELS experiments have been performed on graphene grown on single
crystals of Ru(0001) and Ni(111), by dosing ethylene with the
sample kept at 1050 and 790 K, respectively. For the case of
moir\'{e}-reconstructed \GRRU, periodic nanodomes have
been imaged by STM, as we reported
elsewhere~\cite{borca2010electronic}. HREELS experiments were
carried out by using an electron energy loss spectrometer, whose
energy resolution is 5 meV. The range of primary beam energies
used for investigating the dispersion of low-energy plasmons,
$E_{p}=7-12$ eV, provided the best compromise among surface
sensitivity, high cross-section for mode excitation and $q_{||}$
resolution. As given by
$\hbar q_{||}=\hbar\left(k_{i} \sin\alpha _{i} - k_{s}\sin\alpha
_{s}\right)$, 
the parallel momentum transfer $q_{||}$ depends on $E_{p}$,
$E_{\mathrm{loss}}$, $\alpha _{i}$ and $\alpha _{s}$ according
to~\cite{Rocca19951}:
\begin{equation}\label{Eq:qparallel}
 q_{||}=\frac{\sqrt{2mE_{p}}}{\hbar} \left(\sin\alpha _{i}-\sqrt{1-\frac{{E}_{\mathrm{loss}}}{E_{p}}} \sin\alpha_{s} \right),
\end{equation}
where $E_{\mathrm{loss}}$ is the energy loss, $\alpha _{s}$ is the
electron scattering angle and $k_i$ and $k_s$ are the wave-vectors
of impinging and of scattered electrons, respectively. All
measurements were made at room temperature. Further experimental
details are reported in the Supporting Information (SI).

\begin{figure}[t]
\includegraphics[width=0.99\linewidth]{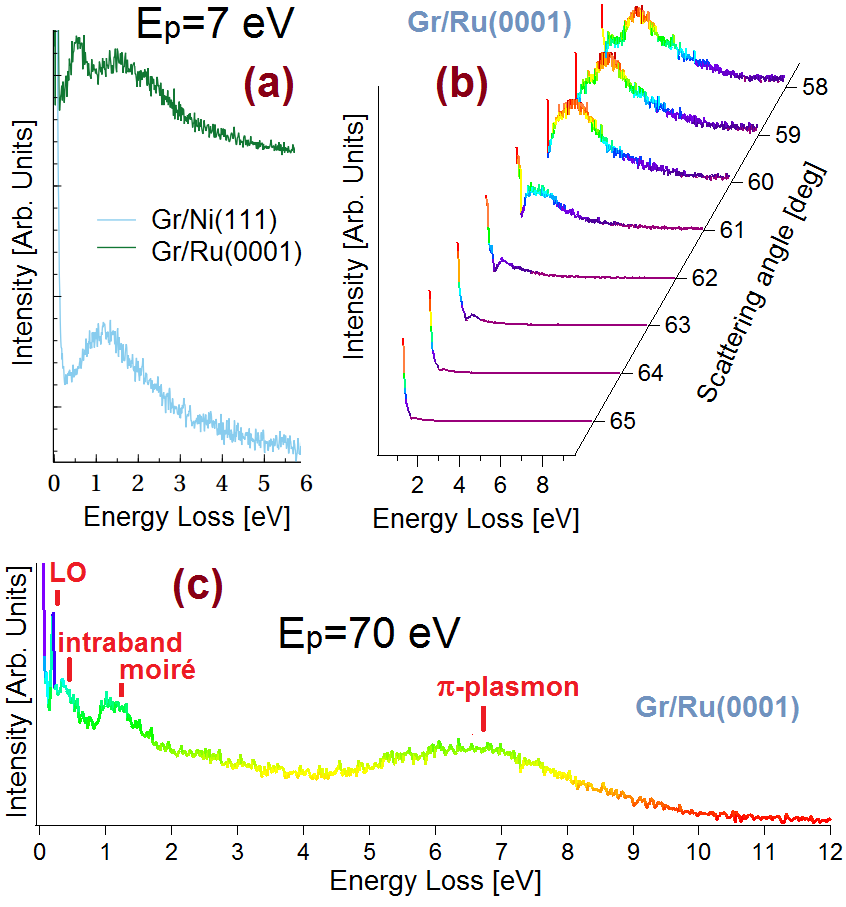}
\caption{ (a) HREELS spectra of graphene (Gr) on Ni(111) and on Ru(0001), acquired with an energy of the primary electron beam $E_p$ of 7 eV; (b) Angle-resolved HREELS spectra for \GRRU, acquired with $E_p=7$ eV and for an incidence angle of 65$^{\circ}$ with respect to the sample normal; (c) Extended HREELS spectrum for \GRRU, measured with $E_p=70$ eV. The longitudinal optical (LO) phonon is also observed.
}
\label{Fig:Fig1}
\end{figure}

Fig. \ref{Fig:Fig1}a shows the HREELS spectra acquired for \GRNI~(blue curve) and \GRRU~(green curve) for a primary
electron beam energy $E_{p}=7$ eV in an off-specular scattering
geometry ($\alpha _{i}=55{^{\circ}}$; $\alpha
_{s}=49{^{\circ}}$). The loss spectrum of \GRNI~is
characterized by a single peak [at an energy of 1.2 eV for $q_{||}=0.12$
\AA$^{-1}$, as obtained from Eq. (\ref{Eq:qparallel})], assigned to the standard intra-band plasmon of doped
graphene (for a review see e.g. \cite{grigorenko2012graphene,S14,politano2014plasmon}). Such a mode corresponds to the collective
excitation of the charge density associated to graphene carriers screened by the Ni substrate \cite{cupo}. The dispersion relation of this mode is that of plasmons in a two-dimensional electron gas (2DEG) \cite{Stern67}
$
\hbar\omega_{pl}(q_{\parallel})\approx \sqrt{\frac{2e^2\mu}{\varepsilon_B}q_{\parallel} + \frac{3}{4} v_F^2q_{\parallel}^2},
$
where $\mu$ is the chemical potential, $\varepsilon_B=(1+\varepsilon_{\rm subs})/2$, $\varepsilon_{\rm subs}$ being the dielectric constant of the substrate and $v_F$ is the Fermi velocity of the carriers. In the long wavelength limit ($q\rightarrow 0$), the main difference is that the velocity of the plasmon mode in \GRNI, defined from the slope of the dispersion relation, is  smaller as compared to isolated graphene, due to the screening effects of the Ni substrate. At large wave-vectors the dispersion of the mode is dominated by the linear term $\sim (3/4)^{1/2}v_Fq_{\parallel}$.

Contrary to the excitation spectrum of graphene on Ni(111) that presents a single peak, our HREELS experiments for
periodically rippled \GRRU, as shown by the green line in Fig. \ref{Fig:Fig1}a,
show two loss features at 0.6 and 1.5 eV, corresponding to
$q_{||}=0.13$ and 0.21 \AA$^{-1}$ in-plane wave-vectors,
respectively. This finding is an evident fingerprint of the deep
difference in the electronic properties of the two
epitaxial-graphene systems. By considering the values of the
above-mentioned physical quantities for \GRRU, we can
assume that the peak at 0.6 eV corresponds to the standard
intra-band plasmon of graphene, which is the counterpart in \GRRU~of the
mode discussed above for \GRNI. Here we are interested on elucidating the
nature of the extra mode observed at $\sim 1.5$ eV, which will be the focus of our work.  For this we have
measured the dispersion relation of plasmonic excitations in
\GRRU, by collecting loss spectra as a function of the
scattering angle at fixed incident angle ($\alpha
_{i}=65{^{\circ}}$) and with $E_{p}=7$ eV. Fig. \ref{Fig:Fig1}b
shows that in nearly specular conditions ($\alpha
_{s}=65{^{\circ}}-62{^{\circ}}$), i.e. for small momenta,  the
intraband plasmon is the unique peak which can be distinguished in
the experimental loss spectrum. At $\alpha _{s}=61{^{\circ}}$, the
extra mode emerges ($q_{||}=0.14$ \AA$^{-1}$) and becomes the
predominant peak of the HREELS spectrum for $\alpha
_{s}=60{^{\circ}}$.

We note that the so-called $\pi$-plasmon mode at $6-7$ eV, originated from inter-band electron-hole transitions between the van Hove singularities of the graphene $\pi$-bands \cite{Eberlein08,Polit,yuan2011excitation,Nelson-NL2014}, cannot be
excited in the experimental conditions at which Fig. \ref{Fig:Fig1}b has been obtained. However, by
increasing the impinging energy $E_{p}$ at 70 eV (Fig. \ref{Fig:Fig1}c), the
$\pi$-plasmon appears in the HREELS spectrum, while the
cross-section for the excitation of low-energy plasmons is reduced
at such a high value of the primary electron beam energy.

\begin{figure}[t]
\includegraphics[width=0.99\linewidth]{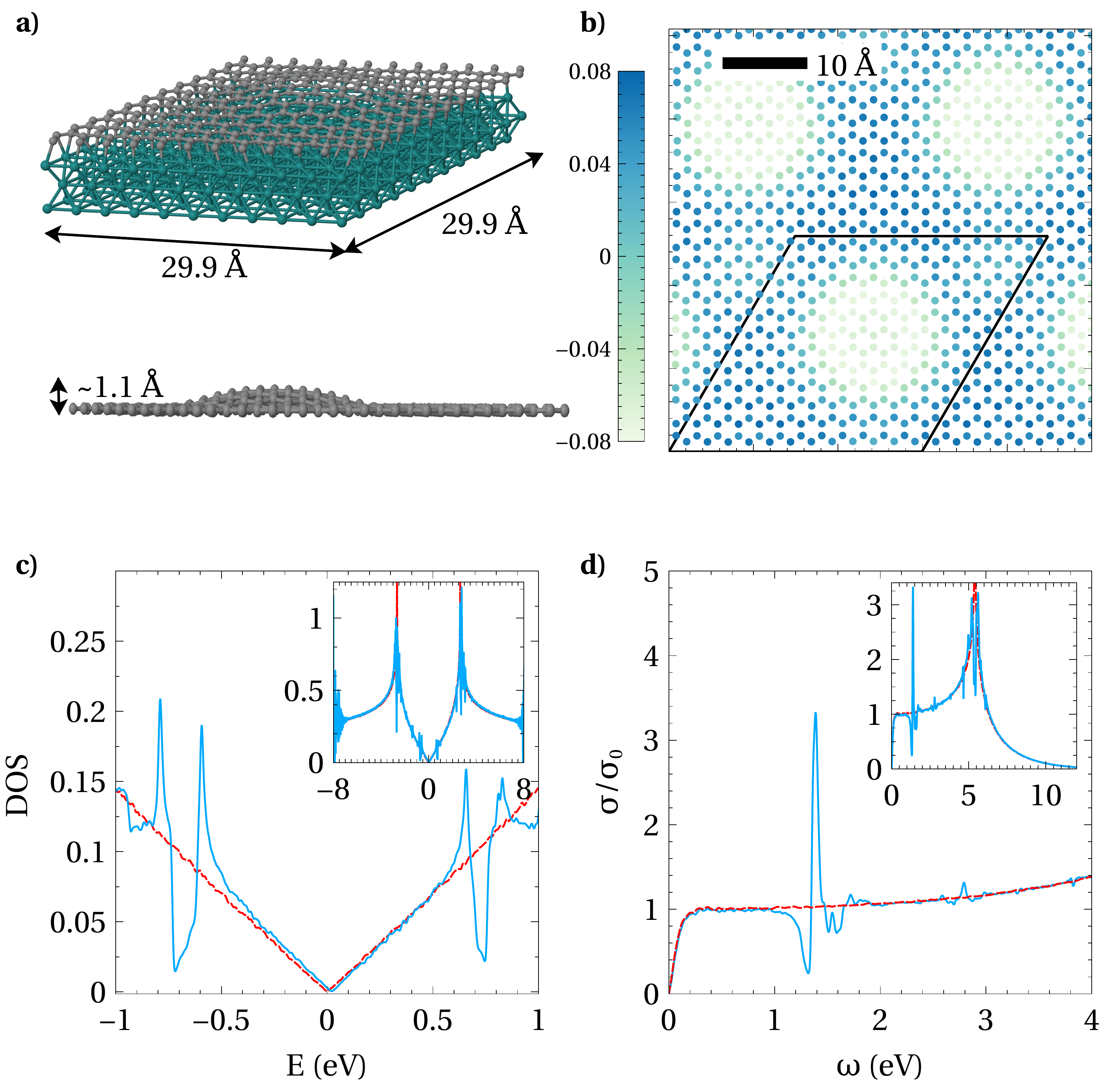}
\caption{ a) The structure of graphene on top of ruthenium as obtained by
DFT calculations. A side view is given on the bottom to indicate the differences in height of the carbon atoms. b) The modified on-site potential in units of eV, with
the DFT unit cell from a) highlighted. c) The density of states for
\GRRU~(blue), which present extra Van Hove singularities associated to the mini bands of the moir\'e supperlattice. For comparison we show the corresponding DOS of a pristine graphene layer (red). The inset shows the DOS for a larger scale. d) The optical conductivity at $\mu=0$ of \GRRU~(blue) as compared to pristine graphene (red), in units of $\protect\sigma _{0}=%
\protect\pi e^{2}/2h$. The inset shows the optical conductivity for larger scales. }
\label{fig:tb-para}
\end{figure}

To establish whether the extra mode could be an effect of the
moir\'{e} superlattice of \GRRU, we have constructed a
TB model of a deformed graphene layer, considering the positions
of the carbon atoms in graphene nanodomes on Ru(0001), as obtained by
DFT calculations (Fig.~\ref{fig:tb-para}a) (see details in Sec. S3
of the SI). We then use this information to modulate the hopping parameters $t_{ij}$ between sites $i$ and
$j$ and on-site potentials $v_{i}$ as function of
interatomic distances of the different carbon
atoms \cite{pereira2009tight,slotman2015effect}.
The modulation of the on-site energy is shown in
Fig.~\ref{fig:tb-para}b, which follows the same periodicity as
the modulation of the atomic coordinates in the moir\'{e} pattern.
By repeating the unit cell shown in Fig.~\ref{fig:tb-para}(a),
we extended the superlattices to $6000\times 6000$ carbon atoms, which is the typical size of our real space TB Hamiltonian matrix. After constructing the TB Hamiltonian for the \GRRU, we
calculate numerically the corresponding electronic density of states (DOS)
by using the TBPM \cite%
{yuan2010modeling,hams2000fast}. Here the DOS is obtained by Fourier
transform of the time-dependent correlation function $\left\langle
\varphi |e^{-iHt}|\varphi \right\rangle $, where $\left\vert
\varphi \right\rangle $ is a random initial wave function. The
time-evolution in the correlation function is performed by using
the Chebychev polynomial algorithm.

The result for the DOS of \GRRU~superlattices is shown
in Fig.~\ref{fig:tb-para}c, which clearly indicates the appearance
of new singularities in the spectrum as compared to pristine
graphene. One could identify some of them according to the
super-periodicity of the moir\'{e} pattern. For example, the extra
singularities around the energy $\approx \pm
0.675$ eV (see inset of Fig. \ref{fig:tb-para}c) follow the
expression $E_{D}=\pm 2\pi \hbar v_{F}/(\sqrt{3}\lambda ) $, where
$\lambda \approx 20.8 a $ is the length of the moir\'{e}'s pattern
in terms of the graphene interatomic distance $a=1.42$~\AA, and $v_{F}=3at/2$
is the Fermi velocity in pristine
graphene~\cite{park2008new,park2008anisotropic}. The emergence of
these extra singularities indicates the presence of flat bands in
the band structure of the moir\'{e} superlattices, and may induce
new strong optical
excitations between the valence and conduction bands. This is indeed what we obtain in our numerical calculations for the optical conductivity shown in Fig.~\ref%
{fig:tb-para}d, which have been obtained by using the Kubo's formula \cite{kubo1957} in the framework of TBPM \cite%
{yuan2010modeling}.
A very sharp optical excitation peak (absent in pristine graphene) appears at the energy
$2\left\vert E_{D}\right\vert \approx 1.35$ eV. At the same time, the
optical conductivity at an energy smaller than $2\left\vert
E_{D}\right\vert $ rapidly drops from $3.3 \sigma _{0}$ to
$0.25 \sigma _{0}$, where $\sigma _{0}=\pi e^{2}/2h$ is the
universal optical conductivity of graphene. Such a phenomenon is qualitatively similar to that obtained in the infrared
optical spectrum of Gr/h-BN heterostructures, due to the quasi-periodic moir\'e superlattice reconstruction, as it has been discussed theoretically
\cite{slotman2015effect} and observed experimentally \cite{Ni2015}. Thus,
periodically rippled graphene on Ru(0001) represents a suitable
platform to modulate the optical properties of graphene.

\begin{figure}[t]
\includegraphics[width=0.45\linewidth]{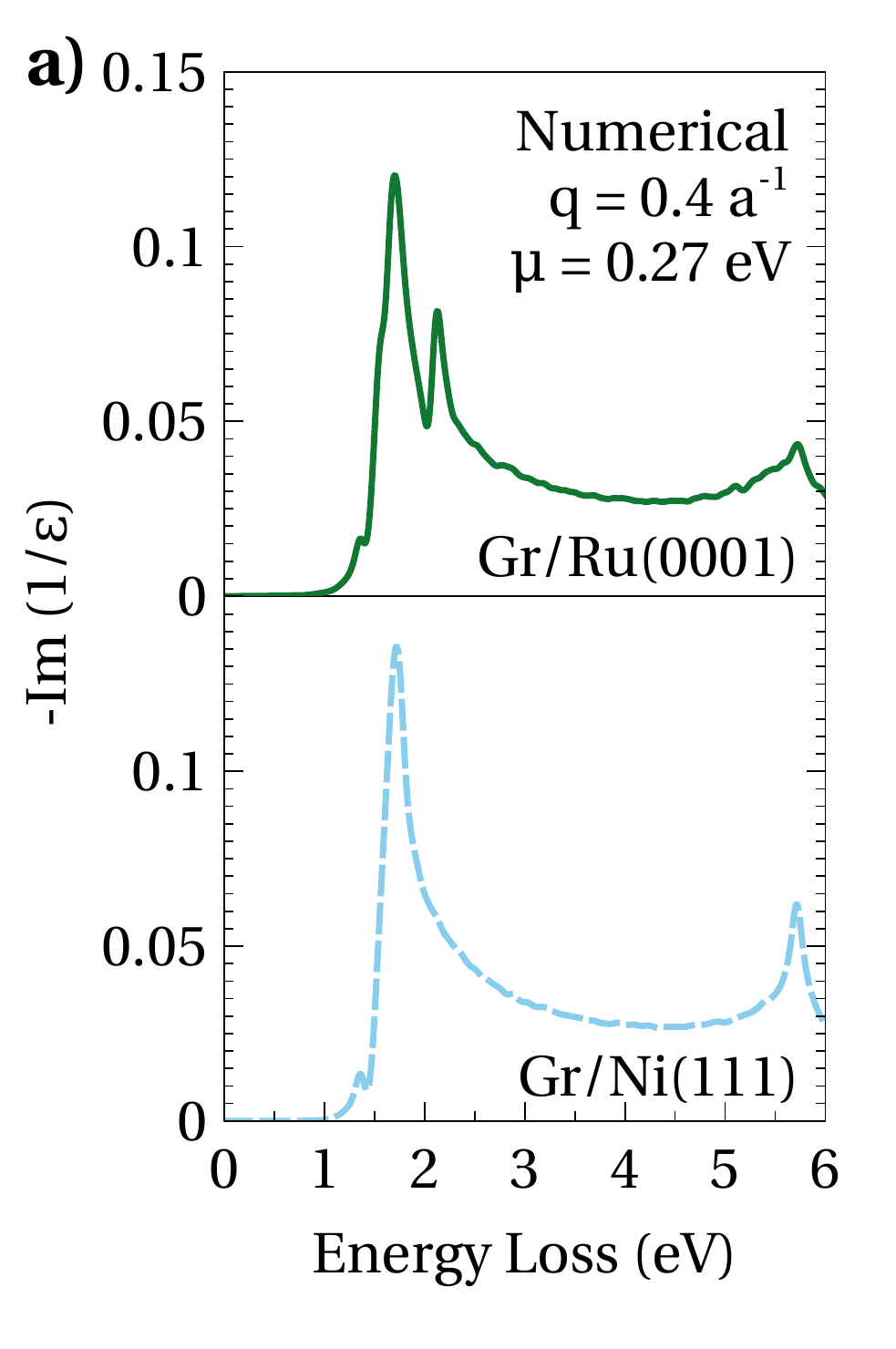}
\includegraphics[width=0.45\linewidth]{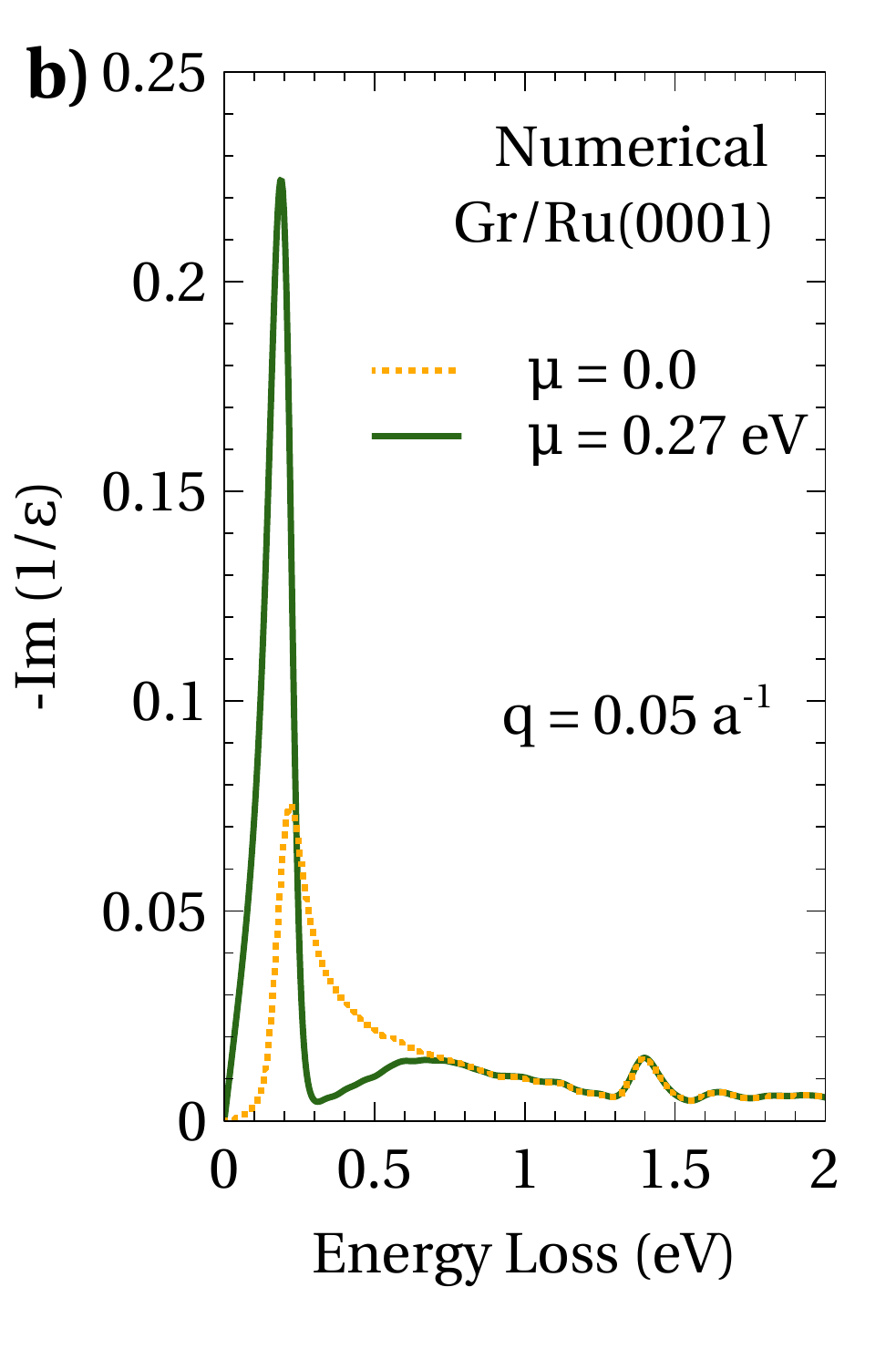}
\caption{a) Numerical results of the energy loss function for  \GRRU~(above) as compared to \GRNI~(below). For \GRRU~an extra peak in the loss function is observed.  b) The loss function for \GRRU~for two different charge doping, namely $\mu = 0$ (dashed yellow) and $0.27$~eV (solid green). Notice that the extra peak is independent of $\mu$.}
\label{Fig3}
\end{figure}

\begin{figure*}[t]
\includegraphics[width=0.32\linewidth]{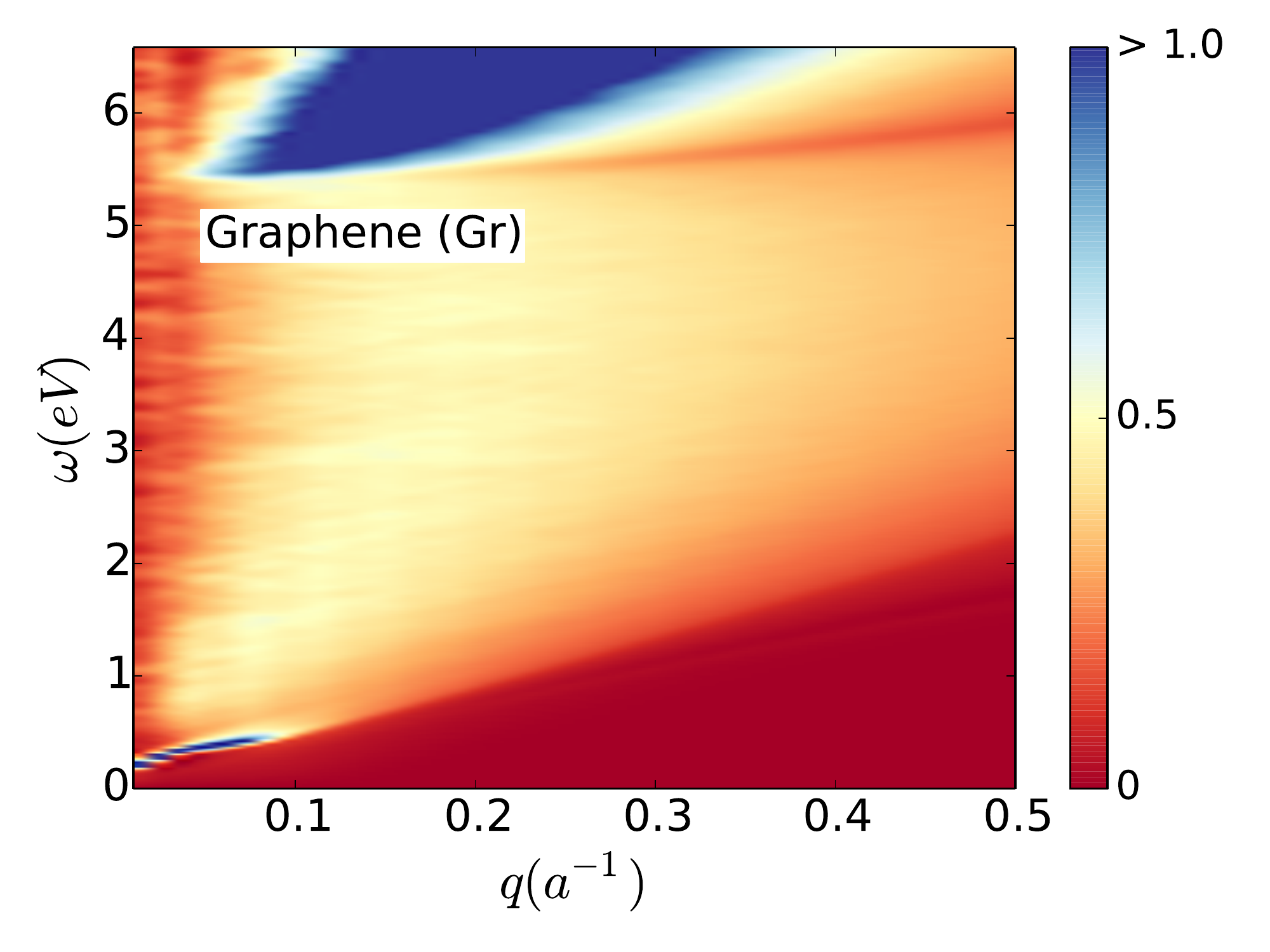}
\includegraphics[width=0.32\linewidth]{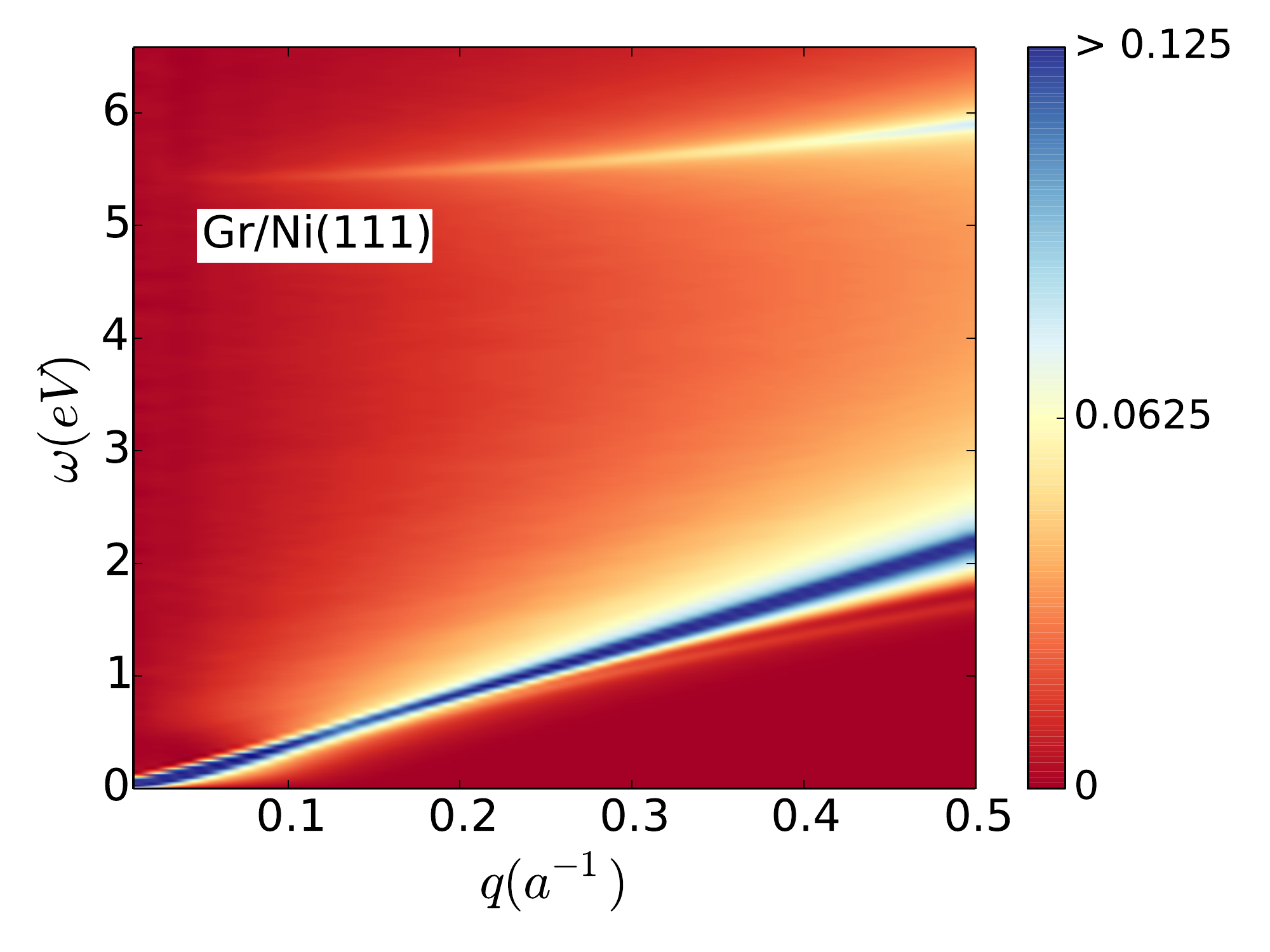}
\includegraphics[width=0.32\linewidth]{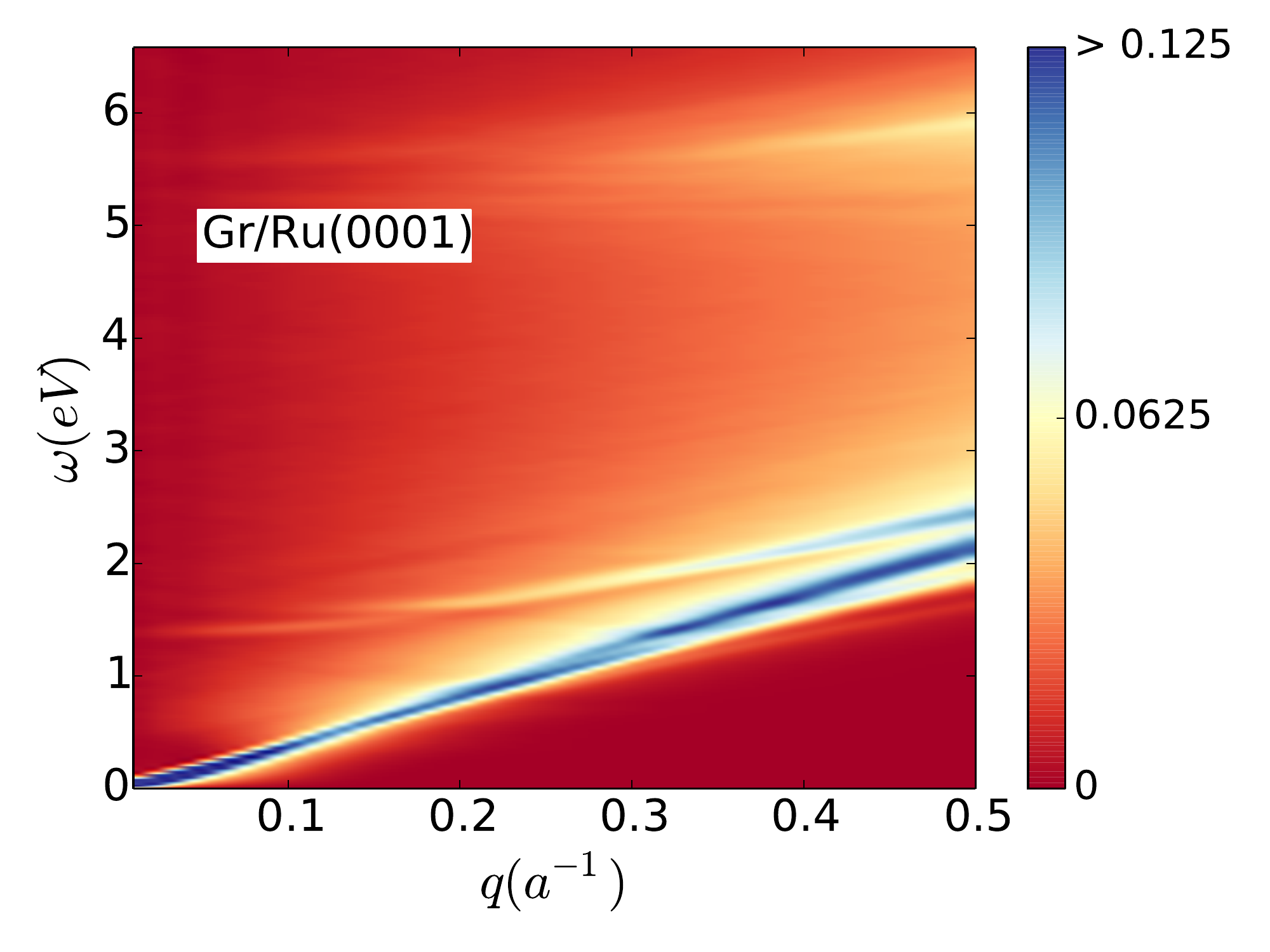}
   \caption{Loss function $- \Im(1/\epsilon(\mathbf{q},\omega))$ calculated in steps of $0.01a^{-1}$ in $q$ for angle $\theta=0^{\circ}$. From left to right: Graphene, \GRNI, \GRRU. For graphene and \GRNI~a flat, unrelaxed layer of graphene is used. For \GRRU~the distortions caused by the substrate are also taken into account. Different dielectric screening constant are used, namely for graphene $\varepsilon_B=1$, and for \GRNI~and for \GRRU, to mimic a metal substrate we have used  $\varepsilon_B=80$.~\cite{choi2006dielectric}
}
\label{fig:idf}
\end{figure*}

In order to compare to the HREELS experiments of Fig. \ref{Fig:Fig1}, we calculate the loss function $-$Im$ \left(1/\epsilon \right)$, in terms of the dielectric function within the random phase approximation (RPA)
$
\epsilon \left(\mathbf{q}, \omega \right) = \mathbf{1} - \frac{2\pi e^2}{q \varepsilon_B} \Pi \left(\mathbf{q}, \omega \right),$
where $\Pi \left(\mathbf{q}, \omega \right)$ is the polarization
function. Our theoretical results for \GRNI~and for
\GRRU~are shown in Fig. \ref{Fig3}. The calculated
loss function $-{\rm Im} \left(1/\epsilon \right)$ of
\GRRU~(Fig. 3a) well reproduces the emerging extra
mode of energy $\sim 1.5$ eV, which is instead absent in flat,
commensurate \GRNI. This fact is in good agreement with our HREELS
measurements, Fig. \ref{Fig:Fig1}. This feature is due to a mode
associated to transitions between electrons in the saddle points
of the mini-bands generated by the moir\'e superlattice.
When devising optoelectronic applications, the behavior of the
modes with respect to modifications of the charge doping should be
taken into account. We have further studied the effect of doping
on the {\it moir\'e} mode. Fig. 3b shows that the extra inter-band
mode is not affected by the carrier concentration, being
present also in a charge-neutral moir\'{e} superlattice on
Ru(0001). The robustness of the moir\'{e} mode toward changes
in electron doping also implies that it would not be influenced by
environmental doping, usually observed in air-exposed graphene
field-effect transistors \cite{Yang2012}. Experimentally, it is not possible to
apply gate voltages in an HREELS apparatus and, thus, we are
unable to probe plasmons in charge-neutral \GRRU.
However, HREELS studies on plasmonic modes in
moir\'{e}-reconstructed, charge-neutral graphene on
Pt$_{3}$Ni(111) reveal that only the moir\'{e} mode exists at
this interface (Section S4, SI).

We notice that, strictly speaking, the moir\'e mode cannot be
considered as {\it fully coherent plasmons}, as it is the case for
the low-energy plasmon with dispersion law given by $
\hbar\omega_{pl}(q_{\parallel})\propto \sqrt{q_{\parallel}} $. For
doped graphene (free-standing or deposited on any substrate), the
intra-band $\sqrt{q_{\parallel}}$ plasmon is undamped above the
threshold $\hbar\omega=v_Fq$ until it enters the inter-band
particle-hole continuum of the spectrum, and then it is damped by
decaying into electron-hole pairs. However, the inter-band {\it
moir\'e} mode that we have found in \GRRU is a mode
which lies inside the continuum of particle-hole excitations:
$-{\rm Im} \Pi(q,\omega)>0$ at the inter-band  energy $\omega$.
Therefore, the mode is damped even in the limit $q\rightarrow 0$
(see also Sec. S7 of the SI). The phenomenology of this mode is
similar to that of the so-called $\pi$-plasmon in graphene, that has been
studied from both experimental \cite{Eberlein08,Polit,Nelson-NL2014} and
theoretical \cite{yuan2011excitation} point of views, and it is
observed also in our present experiments (Fig. \ref{Fig:Fig1}c)
and theoretical calculations (Fig. \ref{fig:idf}) at around $6-7$
eV.

In Fig.~\ref{fig:idf} we compare the corresponding loss functions
$- \Im(1/\epsilon(\mathbf{q},\omega))$ in the
$\omega-q$ plane, calculated with our model in the range of
wave-vectors ($0.01a^{-1}\leq q\leq 0.5a^{-1}$) for momentum
transfer $\mathbf{q}$ along the $\Gamma -$K direction,
corresponding to plasmon wavelengths in the range $\lambda_{pl}
\sim 90 -2$~nm. Plasmon dispersion in the $\Gamma -$M direction
(reported in Figure S6 of the SI) is quite
similar. Both, free-standing graphene and \GRNI~
(Fig.~\ref{fig:idf}a,b), are characterized by two modes: a
low-energy intra-band plasmon (with dispersion relation given by $
\hbar\omega_{pl}(q_{\parallel})\approx
\sqrt{\frac{2e^2\mu}{\varepsilon_B}q_{\parallel} + \frac{3}{4}
v_F^2q_{\parallel}^2} $), and a high-energy inter-band
$\pi$-plasmon \cite{Eberlein08,Polit,yuan2011excitation}. As seen
in Fig. \ref{fig:idf}, the dispersion of the standard low-energy
2D plasmon for all the cases is dominated, at large wave-vectors,
by the linear term $\sim \sqrt{3/4}v_Fq_{\parallel}$, which is
independent of the dielectric constant of the embedding medium.
Periodically rippled \GRRU~(Fig.~\ref{fig:idf}c)
exhibits, besides the above-mentioned modes, an extra mode
dispersing from 1.7 eV at $q=0$ to 2.3 eV at high momenta. As we
have discussed above, we associate this mode to the HREELS peak in
the loss function at this energy (Fig. \ref{Fig:Fig1}). It is
interesting to notice that a careful inspection of
Fig.~\ref{fig:idf}c also indicates the occurrence of
moir\'e-derived splitting of the high-energy inter-band
$\pi$-plasmon. The HREELS spectrum shown in the SI, Fig. S8,
indicates that also such a theoretical prediction well reproduces
the experimental findings.

In conclusion, we have studied the influence of moir\'e
superlattice on the excitation spectrum of Gr/metal
contacts. Whereas \GRNI~shows commensurate lattices
with no relevant effect on the band structure, \GRRU~
leads to a moir\'e superlattice with a reconstruction of the
energy band spectrum of graphene. In particular, new flat mini-bands
are created, which produce the appearance of van Hove
singularities in the electronic DOS. Our HREELS experiments
reveal extra peaks in the loss spectrum of \GRRU,
which are well explained by our theoretical model based on TBPM.
We associate the extra modes to inter-band excitations between the
newly generated van Hove singularities. Our results demonstrate
the importance of the band reconstruction in the excitation
spectrum of graphene on specific substrates, associated to the
moir\'e-derived superperiodic potential.
 The evidence of extra  modes in
graphene quantum dots paves the way toward the control of graphene
plasmonic properties through appropriately engineered periodic
surface patterns. Moreover, the occurrence of an extra mode in
the visible range of the spectrum, robust against environmental doping, opens
novel perspectives for graphene optoelectronics.

\ack
A.P. and G.J.S. contributed equally to this work. AP and GC thank Fabio Vito for technical support in HREELS experiments.
The support by the Netherlands National Computing Facilities foundation (NCF) are
acknowledged. S.Y. and M.I.K. thank financial support from the European
Research Council Advanced Grant program (contract 338957). The research has
also received funding from the European Union Seventh Framework Programme
under Grant Agreement No. 604391 Graphene Flagship.
R.R. acknowledges financial support from MINECO (Spain) through grant FIS2014-58445-JIN.

The Supporting Information is available free of charge on the IOP
Publishing website at DOI:(to be filled) The SI reports
additional information on the TB model; additional experimental
details; the coordinates of the graphene nanodomes on Ru(0001);
Gr/Pt$_{3}$Ni(111); dynamical polarization function and
dispersion relation in the other symmetry directions; splitting in
the $\pi$-plasmon; estimate of the damping of the moir\'e plasmon
mode; loss spectra of clean Ru(0001); and STM visualization of
periodically rippled \GRRU. (PDF)


\section*{References}

\bibliographystyle{iopart-num}
\bibliography{bibliography}

\end{document}